\documentclass[aps,prl,preprint,groupedaddress,showpacs]{revtex4}

\usepackage{graphicx}

\begin{document}


\title{Frozen fronts in cellular flows}


\author{M. E. Schwartz}
\thanks{Current address:  Department of Physics, Columbia University, New York, NY 10027, USA; email:  mes2140@columbia.edu}
\author{T. H. Solomon}
\email[email address: ]{tsolomon@bucknell.edu}
\affiliation{Department of Physics and Astronomy, Bucknell
University, Lewisburg, PA  17837}

\date{\today}

\begin{abstract}
We present experiments on the behavior of reaction fronts in
ordered and disordered cellular flows with imposed winds.  Fronts
in a chain of alternating vortices are found to freeze (pin to the
separatrix) for a wide range of imposed winds that grows
nonlinearly with the characteristic strength of the underlying
vorticity. Experiments in spatially-disordered flows demonstrate
that freezing of fronts is common to cellular flows; furthermore,
it is not dependent on boundary conditions. We therefore
anticipate similar pinning in a wide range of cellular flows and
front-producing systems.
\end{abstract}

\pacs{82.40.Ck, 47.70.Fw, 47.32.C-, 47.54.-r}

\maketitle

Numerous chemical, biological and physical systems are
characterized by two co-existing phases and by the movement of a
front that separates these phases.  Front propagation can be used
to describe a wide variety of dynamical processes, including
natural and industrial chemical processes\cite{grindrod96}, plasma
systems\cite{beule98}, solidification\cite{hurle93}, the spreading
of a disease in a population\cite{kuperman99}, and marine ecology
systems\cite{abraham00}.  The dynamics of front propagation are
well-understood in stagnant fluids\cite{fisher,kpp}; an issue of
significant current interest\cite{tel2005} is how fluid flows
affect the motion of fronts. If a uniform ``wind'' opposing the
front is applied to an otherwise motionless fluid, the front
simply propagates at its reaction-diffusion (no flow) velocity
minus the wind speed.  If the same wind is applied to a fluid with
an underlying cellular flow, however, the behavior is dramatically
different.

In this Letter, we present experiments showing that cellular flows
can freeze the motion of a reaction front against an imposed wind.
We study this phenomenon as a function of the strength of the
vortex flow.  Fronts produced by the ruthenium-catalyzed excitable
Belousov-Zhabotinsky (BZ)
reaction\cite{winfree72,showalter80,field85,tinsley05} are studied
in both ordered and disordered vortex flows. The flows studied are
all time-independent; nevertheless, the experimental results have
implications for time-dependent flows as well, since a moving
vortex in a time-dependent flow can be viewed in a co-moving
reference frame as a stationary vortex with a time-dependent
imposed wind. We therefore expect front-freezing to be relevant
for a wide range of cellular flows.

In the absence of fluid flows, a front propagates with a
reaction-diffusion (RD) velocity given by the well-known FKPP
result\cite{fisher,kpp} $v_0=2\sqrt{D/\tau}$, where $D$ is the
molecular diffusivity and $\tau$ is the reaction timescale. In the
past 10 years, there has been growing interest in the effects of
fluid advection on front propagation. Theoretical\cite{edwards02}
and experimental\cite{leconte03} studies have shown that a front
propagating against a simple pipe flow moves with velocity $v_0$
independent of the imposed flow due to no-slip boundary conditions
at the walls.  No-slip conditions also play a key role in
explaining anomalous front velocities in flows through a porous
medium\cite{kaern02}.   Other theoretical studies of front
propagation in advection-reaction-diffusion (ARD) systems include
predictions of fractal fronts in open flows with chaotic
mixing\cite{neufeld01,toroczkai98}.  Recent
theories\cite{mancinelli02,diego03,brockmann07} have extended the
standard FKPP theory to general ARD systems; however, these
approaches do not account explicitly for the effects of cellular
flows.  The importance of cellular flow structures, for example,
is in evidence in time-periodic vortex chains, which cause fronts
to mode-lock to the external forcing\cite{cencini03,paoletti05}. A
mode-locked front propagates an integer number of vortex pairs in
an integer number of oscillatory drive periods, inconsistent with
straightforward FKPP approaches.  Our experiments indicate the
fundamental importance of coherent vortices in any general
description of ARD dynamics.

\begin{figure}
\begin{center}
\includegraphics[width=7.0cm]{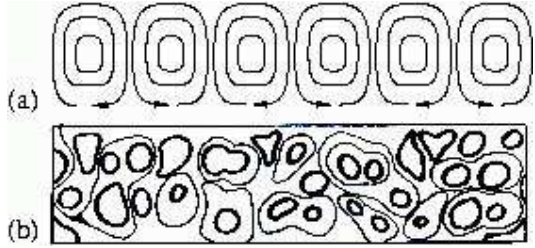}
\caption[Schematics.]{Schematics of the fluid flow for $\epsilon =
0$ (no flow).  (a) Ordered chain of counterrotating vortices. The
maximum vortex velocity $U$ is measured at the separatrix.  (b)
Two-dimensional disordered array.} \label{fig1}
\end{center}
\end{figure}

Two flows are used for most of these experiments
(Fig.~\ref{fig1}): a chain of 20 counter-rotating vortices and a
random vortex flow, both created using magnetohydrodynamic
forcing\cite{paoletti05b} and confined to an annulus bounded by
Plexiglass rings of radii 6.1 and 8.3 cm.  A radial current
passing through a 2 mm-thick electrolytic solution interacts with
a magnetic field produced by Nd-Fe-B magnets below the fluid.  Two
concentric rings of 3/4"-diameter magnets with alternating
polarity are used for the ordered vortex chain and a disordered
pattern of 1/4" magnets is used for the random flow. The magnets
are mounted on a motor that rotates at a constant rate, thus
moving the vortices.  In a reference frame moving with the
magnets, the vortices are stationary and there is a constant,
uniform wind with a velocity $W$ equal to the drift velocity of
the magnets.

The electrolytic solution is composed of the chemicals for the
excitable Ru-catalyzed BZ reaction\cite{tinsley05}. In this
reaction, orange Ru$^{2+}$ ions are oxidized to a green Ru$^{3+}$
state, forming a propagating pulse-like front which is imaged
using a 12-bit CCD camera with a red interference filter. The
Ru-catalyzed BZ reaction is also inhibited by blue-green light. To
confine the reaction to the area of interest, we use a video
projector to shine blue light everywhere except in the
annulus\cite{paoletti05b}. We also illuminate a small section of
the annulus (1-2 vortices) with blue light to control the
direction of front propagation. The reaction is triggered with a
silver wire on one side of this blinding region; the front cannot
propagate through the blinding region and therefore only
propagates in one direction.  The enhanced images are digitally
``de-curled'' into a linear chain and shifted into a reference
frame moving with the vortices for analysis.

\begin{figure}
\begin{center}
\includegraphics[width=9.0cm]{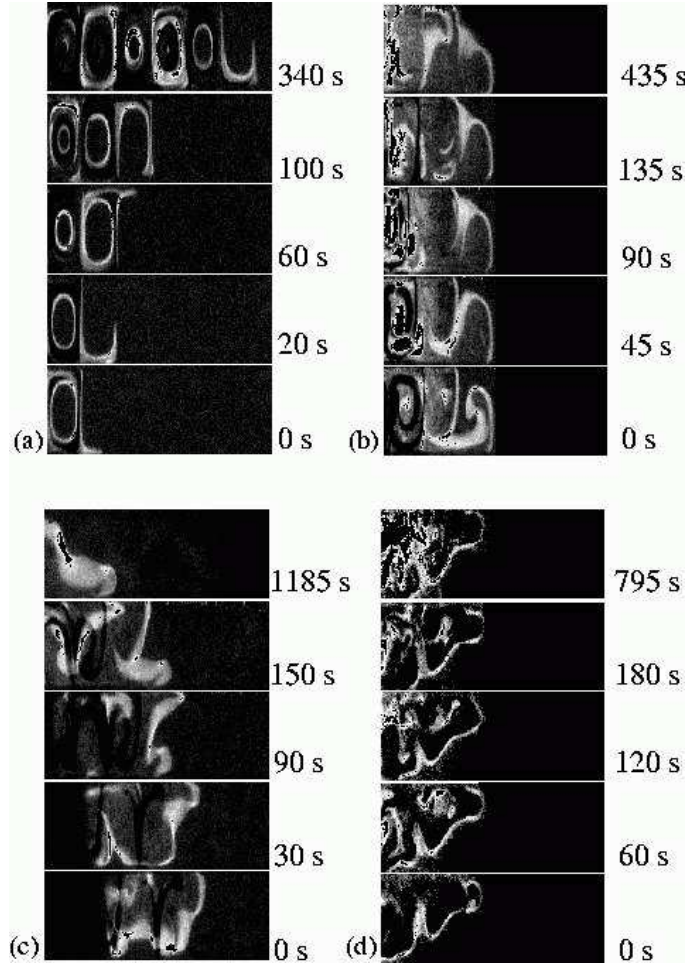}
\caption[Sequences]{Sequences of de-curled images, showing regimes
of front propagation.  In all cases, the imposed wind blows from
right to left.  (a) Forward-propagating front; $\epsilon = 0$ and
$\mu = 40$. The front is advected around the vortex and burns
across the separatrix in order to propagate forward.  (b) Frozen
front; $\epsilon = 2.6$ and $\mu = 12$.  The wind prevents the
reaction from burning across the separatrix.  (c)
Backward-propagating front; $\epsilon = 8.6$ and $\mu = 12$.  The
front is initially in the leading vortex but is blown backwards by
the wind.  (d) Frozen front for a narrow random array of vortices;
$\epsilon = 4.0$ and $\mu = 12$.} \label{fig2}
\end{center}
\end{figure}

The front velocity $v_f$, wind speed $W$, and maximum vortex
velocity $U$ are all scaled by the RD velocity: $\nu = v_f/v_0$,
$\epsilon = W/v_0$ and $\mu = U/v_0$.  In the limit $\mu
\rightarrow 0$, the addition of a uniform wind is the equivalent
of a Galilean transformation; in a reference frame moving with the
wind, there is no flow. Consequently, for $\mu = 0$
\begin{equation}
\nu = 1 - \epsilon
\label{eqn1}
\end{equation}
In this limit, a front is ``frozen'' (i.e., $\nu = 0$) only if $W$
is precisely equal to $v_0$, i.e. if $\epsilon = 1$.

\begin{figure}
\begin{center}
\includegraphics[width=12.0cm]{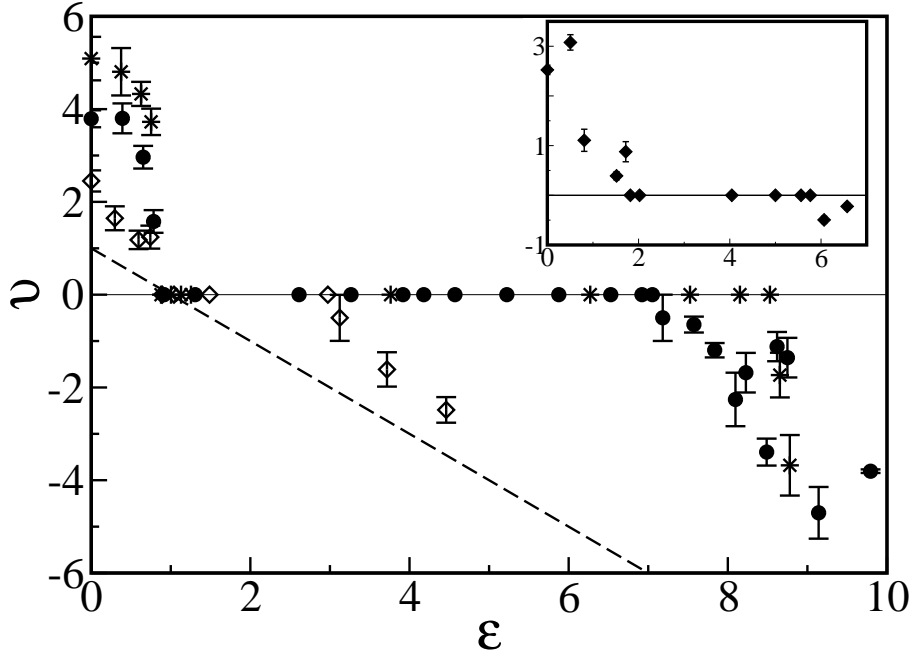}
\caption[vf]{Front velocities in the face of opposing wind.  The
data shown is for $\mu = 4$, $12$ and $40$ (open diamonds, filled
circles and stars respectively) along with the theoretical limit
(Eq.~\ref{eqn1}) for $\mu = 0$ (dashed line).  The inset shows the
same data for a narrow  random chain of vortices at $\mu = 12$.}
\label{fig3}
\end{center}
\end{figure}

The addition of underlying vortex structures ($\mu \ne 0$) has a
significant effect on front behavior.  Sequences of images
(Fig.~\ref{fig2}) show the regimes of front propagation in this
case. In an ordered vortex flow with small $\epsilon$
(Fig.~\ref{fig2}a), the front propagates forward against the wind;
it is advected around each vortex and then burns across the
separatrix from one vortex to the next.  The front freezes for
intermediate wind speeds (Fig.~\ref{fig2}b):  the wind prevents
the front from burning across the separatrix into the next vortex.
However, the wind does not blow the front backwards even though
$\epsilon$ significantly exceeds 1; instead, the front circles in
the leading vortex.  In the co-moving reference frame of
Fig.~\ref{fig2}, the front is {\it frozen}; in any other reference
frame, the front is {\it pinned} to the motion of the leading
vortex. For large enough $\epsilon$ (Fig.~\ref{fig2}c), the front
is pushed back by the wind.

Front velocities in the co-moving reference frame are plotted as a
function of the wind speed in Fig.~\ref{fig3} for three values of
$\mu$, along with the prediction from Eq.~\ref{eqn1} for the $\mu
= 0$ limit. The most salient feature of Fig.~\ref{fig3} is the
plateau, clearly visible for all three values of $\mu$, where the
front velocity $\nu = 0$.  The width of the frozen-front plateau
decreases with decreasing $\mu$ and the velocities approach the
theoretical limit as $\mu$ approaches 0. Also note that advection
due to the vortices enhances the front speeds when $\epsilon < 1$,
consistent with previous studies of the $\epsilon = 0$
limit\cite{abel01,paoletti05b,pocheau06}.

\begin{figure}
\begin{center}
\includegraphics[width=12.0cm]{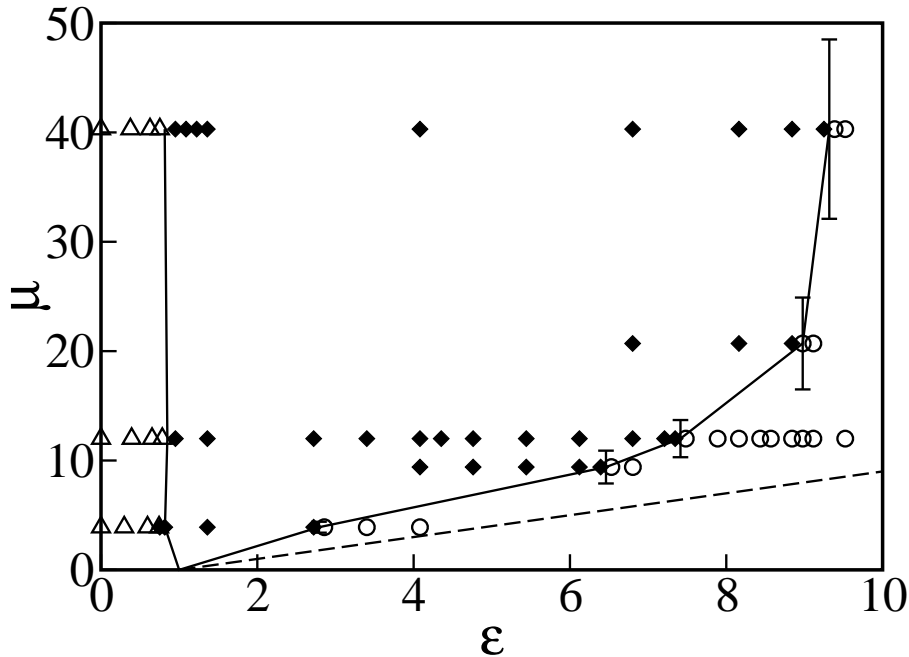}
\caption[parameter space diagram]{Parameter-space diagram, showing
the increase in the range of the frozen front regime with
increasing $\mu$.  Open triangles denote forward propagation of
the front against the wind, closed diamonds represent pinned
fronts, and open circles denote backward (downwind) front
propagation. The solid line shows the boundary between pinned and
unpinned fronts. The error bars show the uncertainty in $\mu$,
which is caused chiefly by uncertainty in $U$ as it increases. The
dashed line is at $\epsilon = \mu + 1$, the normalized sum of the
RD and advective velocities.} \label{fig4}
\end{center}
\end{figure}

The variation with $\mu$ of the width of the pinned-front regime
can be seen in a parameter-space diagram (Fig.~\ref{fig4}). The
minimum wind speed to achieve frozen fronts is (within error) the
RD front velocity: $\epsilon = 1$.  This can be understood by
considering the behavior near the separatrices, where the wind is
perpendicular to the underlying vortex flow and where forward
propagation is not aided by advection.  If $\epsilon$ exceeds 1,
the wind is stronger than the forward-burning RD velocity, and the
front stalls at the separatrix.  The endpoint of the plateau is
significantly below $\mu + 1$ (i.e. the sum of the RD velocity and
the advective velocity), and it diverges from this line in a
nonlinear fashion. We are currently investigating this nonlinear
behavior, particularly in light of secondary flows\cite{solomon03}
and no-slip boundary conditions in the experiments.

Frozen fronts are also observed for random vortex flows, as shown
in Fig.~\ref{fig2}d and the inset of Fig.~\ref{fig3}. The freezing
mechanism remains the same -- the front is pinned at the (more
convoluted) separatrices. Several features, however, are different
for the random flow.  First, the front velocity does not decrease
smoothly as $\epsilon$ approaches 1, due to the random nature of
the flow -- different sections of the flow have different
characteristic advective speeds. Second, the minimum $\epsilon$
necessary for frozen fronts is significantly greater than 1. The
separatrices are not typically perpendicular to the wind for
random flows; as a result, higher $W$ is required for the
perpendicular component of the wind to exceed $v_o$ and thus
prevent the front from burning across a separatrix.

\begin{figure}
\begin{center}
\includegraphics[width=12.0cm]{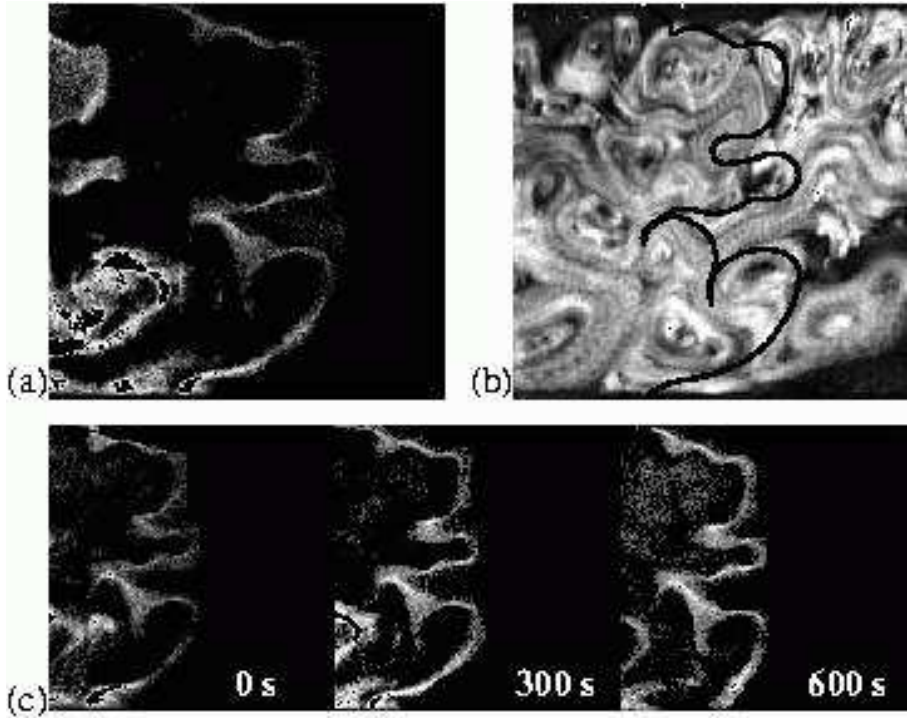}
\caption[]{Frozen front in a wide two-dimensional random array of
vortices.  The images are de-curled with the inner and outer radii
at the bottom and top of the image, respectively.  (a) Image of
the frozen reaction front.  The normalized wind $\epsilon$ ranges
from 1.8 to 5.8 from the bottom to the top of each image.  (b)
Streak photograph of the underlying vortex flow with $\epsilon =
0$. The front from (a) (black curve) is superimposed on the
section of the flow where it freezes. (c) Snapshots of frozen
front.} \label{fig5}
\end{center}
\end{figure}

Freezing of fronts is not restricted to flows with a limited
number of vortices, nor is it dependent on no-slip boundary
conditions in a confined geometry.  A frozen front is shown in
Fig.~\ref{fig5} for a significantly wider annulus (inner and outer
radii 2.6 and 8.3 cm), along with a streak photograph showing the
underlying disordered pattern of vortices.  Due to the annular
geometry, $W$ grows by a factor of 3 between the inner and outer
edges; despite this increase in $W$ fronts freeze across the
entire annulus, a further indication of the robustness of this
phenomenon.  As seen in Fig.~\ref{fig5}b, the frozen front
generally follows the separatrices, although it is pushed back
slightly from the edge of the vortices at larger radii due to the
increased wind.

Our experiments indicate that pinning of reaction fronts should
occur in a wide range of steady vortex flows, regardless of the
spatial pattern of the vortices.  Furthermore, these results help
interpret and predict front behavior in time  {\it dependent}
flows. flows.  We can use front pinning, for example, to explain
mode-locking in an oscillating vortex
flow\cite{cencini03,paoletti05}. In a co-moving (i.e. oscillating)
reference frame, the vortices are stationary in the presence of an
oscillating wind.  For sufficiently large oscillatory velocity,
the wind pins the front to a vortex during a significant fraction
of each oscillation period.  Thus, the front can only propagate
during a well-defined segment of each period, tying the front
propagation speed to the oscillation frequency.

Moving vortices in other time-dependent flows can be expected to
pin reaction fronts similarly, if only temporarily, in a manner
that fundamentally alters the speed and method of front
propagation.  Consider the case of a vortex and a front moving in
the same direction.  If the vortex passes through the front with a
sufficient speed, it will pin and drag the front forward.  (In the
reference frame of the vortex, this is equivalent to the
frozen-front case.)  Even if the vortex later slows down or speeds
up such that it can no longer hold the front, it will have already
significantly altered the shape and location of the front.  These
ideas should extend even to turbulent flows, which are often
characterized by both transient and long-lived coherent
vortices\cite{sommeria88,marcus88}. Ultimately, any general theory
of front propagation in ARD systems will have to account for
pinning of fronts by vortices in the flow.

\begin{acknowledgments}
This work was supported by the US National Science Foundation
(grants DMR-0404961, DMR-0703635 and REU-0552790).
\end{acknowledgments}



\clearpage

\end{document}